\documentclass[%
 reprint,superscriptaddress,
 amsmath,amssymb,
 aps,
]{revtex4-2}

\usepackage{array}
\usepackage{float}
\usepackage{graphicx} 
\usepackage{amsmath,amsfonts,amssymb,graphicx}
\usepackage[english]{babel}
\usepackage[utf8]{inputenc}
\usepackage[T1]{fontenc}
\usepackage{bbold}
\usepackage{caption,subcaption}
\usepackage{color}
\usepackage{algorithmic}
\usepackage{hyperref}
\usepackage[ruled,lined]{algorithm2e}
\usepackage{blindtext}
\usepackage[color=yellow]{todonotes}
\usepackage{booktabs}
\usepackage{multirow}
\usepackage{colortbl} 
\usepackage{xcolor}
\usepackage{tikz}
\usetikzlibrary{quantikz2}
\usepackage{dsfont}
\usepackage{wrapfig,tcolorbox,cleveref}

\newtcolorbox[auto counter,crefname={box}{boxes}]{pabox}[2][]{%
title=Box 1 Timeline,
colback=white,
colframe=black,
colbacktitle=white,
coltitle=black,
fonttitle=\bfseries,
title=Box ~\thetcbcounter. #2, label={#1}}

\makeatletter
\newcommand*{\balancecolsandclearpage}{%
  \close@column@grid
  \cleardoublepage
  \twocolumngrid
}
\makeatother

\begin{document}


\title[]{Efficient algorithm for fidelity estimation of two quantum states}

\author{Anumita Mukhopadhyay}
\email{anumitamukherjee455@gmail.com}
\affiliation{Center for Quantum Engineering, Research and Education (CQuERE), TCG CREST, Salt Lake, Sector 5, Kolkata 700091, India.}
\affiliation{Academy of Scientific and Innovative Research (AcSIR), Ghaziabad- 201002, India.}
\author{Shibdas Roy}
\email{roy.shibdas@gmail.com}
\affiliation{Center for Quantum Engineering, Research and Education (CQuERE), TCG CREST, Salt Lake, Sector 5, Kolkata 700091, India.}
\affiliation{Academy of Scientific and Innovative Research (AcSIR), Ghaziabad- 201002, India.}
\author{Arun Kumar Pati}
\email{patiqubit@gmail.com}
\affiliation{Centre for Quantum Technology,
KIIT University, Bhubaneswar, Odisha, India.}

\begin{abstract}
The fidelity estimation between two quantum states is crucial for quantum computation and information science. However, an efficacious method for this, especially for mixed states and higher-dimensional density matrices, remains elusive. While there are many existing algorithms on computing the fidelity between two pure states, there is not much work on how to obtain the fidelity between two mixed states. Here, an efficient quantum algorithm for the fidelity estimation is proposed, based primarily on the density matrix exponentiation and interferometeric scheme for mixed states, with a time complexity of $O(\kappa^2N^2/\epsilon^7)$, where $N$ is the system size, $\kappa$ is the larger of the condition number of the density matrices and $\epsilon$ is a precision error. This algorithm may serve as a resource-efficient technique to deduce fidelity of any two (pure or mixed) unknown or known quantum states, when the density matrices of the quantum states commute with each other.
\end{abstract}

\maketitle

\section{Introduction}

In quantum information science,  the notion of fidelity estimation between two density operators is a cornerstone serving as a crucial metric for quantifying the ``closeness'' of quantum states. Its importance stems from its direct relevance to various fundamental and applied aspects of quantum mechanics. Firstly, in quantum state tomography, where the goal is to reconstruct an unknown quantum state, fidelity provides a robust measure of how accurately the reconstructed state matches the true state, thereby validating the experimental procedure. Secondly, in quantum computing and quantum simulation, fidelity is indispensable for assessing the performance of quantum gates and algorithms. By comparing the experimentally prepared or evolved state with the theoretically ideal target state, one can quantify the error rates and evaluate the effectiveness of error correction schemes \cite{qec,qec2}. High fidelity is a prerequisite for achieving fault-tolerant quantum computation. Thirdly, in quantum communication, fidelity helps to determine to what extent the quality of transmitted quantum information is maintained across noisy channels \cite{PhysRevA.83.053420,qcc}. Low fidelity indicates significant decoherence or loss of information. Beyond these, fidelity plays a vital role in quantum metrology for evaluating the precision of quantum sensors \cite{metrology}, in quantum thermodynamics for analyzing the efficiency of quantum engines, in supervised and unsupervised learning schemes in quantum machine learning \cite{qml} and in fundamental quantum mechanics for understanding concepts like entanglement and quantum correlations. Its ability to provide a single, intuitive value for state similarity makes fidelity estimation an indispensable tool for benchmarking \cite{engg}, validating, and advancing quantum technologies and our understanding of quantum phenomena. To this end, fidelity, a metric that captures the ``closeness'' between two quantum states represented by density matrices, $\rho_1$ and $\rho_2$, serves as a critical tool.
We define fidelity between density matrices of two quantum states, say, $\rho_1$ and $\rho_2$ by $F(\rho_1,\rho_2)$, as given below \cite{jozsa,transition,qcode,NC}: 
\begin{eqnarray}\label{eq: Fidelity}
    F(\rho_1,\rho_2) &=& \left[{\rm Tr}\sqrt{\sqrt{\rho_1}\rho_2\sqrt{\rho_1}}\right]\\
    &=& [{\rm Tr}|\sqrt{\rho_1 \rho_2}|]
\end{eqnarray}
Eqn. (\ref{eq: Fidelity}) can be simplified as $F(\rho_1,\rho_2)={\rm Tr}\sqrt{\rho_1 \rho_2}$ when $\rho_1$ and $\rho_2$ commute with each other. 
Here, $\rho= \ket{\psi_{\rho_1}}\bra{\psi_{\rho_1}}$ and $\rho_2= \ket{\psi_{\rho_2}}\bra{\psi_{\rho_2}}$, if $\rho_1$ and $\rho_2$ are pure states. Thus, for pure states, $F(\rho_1,\rho_2)=|\braket{\psi_{\rho_1}}{\psi_{\rho_2}}|^2$, where $\rho_1$ and $\rho_2$ are positive semi definite. That is, for pure states, $F$ is the probability of finding the state $\ket{\psi_{\rho_1}}$ when measuring $\ket{\psi_{\rho_2}}$ in a basis containing $\ket{\psi_{\rho_1}}$. Note that $F(\rho_1, \rho_2)=0$, if $\ket{\psi_{\rho_1}}$ and $\ket{\psi_{\rho_2}}$ are orthogonal, in which case they are perfectly distinguishable. By contrast, if $\ket{\psi_{\rho_1}}=\ket{\psi_{\rho_2}}$, we have $F(\rho_1,\rho_2)=1$. Fidelity is a significant measure as it is well-defined for both pure and mixed states. $F(\rho_1,\rho_2)$ is symmetric under the exchange of $\rho_1$ and $\rho_2$. This is also an invariant quantity under unitary operation. 

In order to find the fidelity of two unknown quantum states, we usually need to characterize the states. This can be done by quantum state tomography \cite{NC,spintomo,qtomo}, which suffers from an exponential requirement of resources, particularly for a higher-dimensional system. Later, some more efficient procedures of quantum state tomography were proposed \cite{Dgross,efficient_tomo,effqtomo}. After the unknown states are estimated by tomography, fidelity is measured by matrix algebra. To bypass the requirement of computationally expensive processes, different groups have proposed various algorithms to estimate fidelity between two pure quantum states \cite {purefid1,purefid,altfid}. Some of the research also propose direct measure of fidelity for pure states \cite{directfidelity,directfidelity2}. A recent work suggests direct fidelity estimation for one pure and one mixed state using classical shadow tomography and quantum amplitude estimation \cite{directfidelity3}. Some algorithms are also available to estimate the fidelity of two general quantum states using variational quantum algorithms \cite{varfid}. There are certain algorithms for finding the fidelity of two mixed quantum states also \cite{fidelity}, that are efficient for lower rank of density matrices. Computing Uhlmann fidelity when the description of density matrices are available explicitly is also shown in \cite{baldwin}.

In this work, we design a new algorithm to find the fidelity of two unknown general (pure and mixed) quantum states which commute with each other. 
Instead of finding the full description of the state $\rho_1$ and $\rho_2$, we use the Lloyd-Mohseni-Rebentrost algorithm (LMR) technique \cite{LMR} to exponentiate the density matrices, to be used as controlled unitaries in improved quantum phase estimation (IQPE) \cite{HHL} to estimate the eigenvalues of $\rho_1$ and $\rho_2$, followed by controlled rotation and post-selection, to get the normalised square roots of the density matrices. Subsequently, we use the interference of the mixed state scheme, which utilizes a Mach-Zehnder interferometer to determine the geometric phase of the mixed state evolving under unitary transformation \cite{akp,KPP}. We use this to obtain a quantity involving the product $\rho_1\rho_2$, which we require for the purpose of computing fidelity as given in (\ref{eq: Fidelity}). By circumventing full state tomography, our protocol appears as a resource efficient estimation.

Commuting density matrices do appear in quantum information science, though typical scenarios often involve non-commuting states. Here are some important settings where commuting density matrices naturally arise: (i) Classical Mixtures: When quantum states represent probabilistic mixtures of orthogonal basis states (such as diagonal density matrices in the computational basis), all such density matrices commute.
These are common in discussions of decoherence, quantum-to-classical transitions, and noise models that fully dephase in one basis. (ii) Quantum Channels with Classical Noise:
In quantum channels that induce classical (phase-damping or dephasing) noise, output states from different inputs are often diagonal and thus mutually commuting. For a complete dephasing channel, any two output density matrices remain diagonal in the same basis and therefore commute.
(iii) Projective Measurements in a Fixed Basis:
After projective measurements (especially in the computational basis), the resulting post-measurement states are diagonal and mutually commuting, representing classical outcomes. This is common in quantum tomography and state estimation with projective measurements.
Also, in scenarios where measurements are performed with observables that commute, the corresponding projectors (and hence, the post-measurement density matrices) also commute.
(iv) Quantum Statistical Mixtures from Macroscopic Observables:
When two density matrices describe populations that are only distinguished by macroscopic, commuting observables (like energy in thermal states within the same Hamiltonian with non-degenerate spectrum), the matrices commute. Hence, developing a new algorithm to efficiently estimate fidelity between two commuting density matrices is worthwhile and meaningful in quantum information theory.

Our paper is organised as follows. In Section \ref{2}, we have discussed the mathematical background of the problem and methodology to solve it. In Section \ref{3}, we lay down the consolidated steps of the algorithm. In Section \ref{4}, we have calculated the complexity of our scheme, following the steps of Section \ref{3}. In Section \ref{5} we provide a numerical example to verify our algorithm using Mach-Zehnder interferometric scheme. We discuss a few key insights about our algorithm in Section \ref{6} and conclude in Section \ref{7}.

\section{Method}\label{2}
The initial stage of our scheme involves creating a quantum state, that is the normalized square root of an unknown quantum state, designated as $\rho$. Since our goal is to ascertain the fidelity ${\rm Tr}\sqrt{\rho_1\rho_2}$ between two unknown commuting quantum states, $\rho_1$ and $\rho_2$, we accomplish this through a combination of Density Matrix Exponentiation (DME) using Lloyd-Mohseni-Rebentrost algorithm (LMR) and Improved Quantum Phase Estimation (IQPE), followed by a controlled rotation of an ancilla qubit and a post-selection of the state conditioned on the ancilla qubit. As a result, we obtain the state $\sqrt{\rho}/{\rm Tr}(\sqrt{\rho}$). We additionally can estimate the value of ${\rm Tr}(\sqrt{\rho})$ by performing another DME on the ancilla qubit, followed by IQPE for the eigenstate $|1\rangle$. To obtain the fidelity subsequently, we will leverage a Mach-Zehnder Interferometer setup, shown in Figure \ref{fig:mzi_circuit}, to precisely measure the geometric phase, and relate it to the expression of desired fidelity. If we send a density operator $\rho'$ through the intereference setup, the intensity observed in the Mach-Zehnder Interferometer is given by \cite{akp}
\begin{eqnarray}
   I \propto [1+{\rm Tr}(\rho^\prime U)\cos(\phi + \arg ({\rm Tr}(\rho^\prime U)))],
\end{eqnarray}
where $\arg ({\rm Tr}(\rho^\prime U))$ gives the phase. Here, we will take $\rho^\prime =\sqrt{\rho_1}/{\rm Tr}\sqrt{\rho_1}$, and $U= e^{it_2 \sqrt{\rho_2}/{\rm Tr}\sqrt{\rho_2}}$ used as the controlled unitary, so that we have $U=\mathbb{1}-i\tau \frac{\sqrt{\rho_2}}{{\rm Tr}\sqrt{\rho_2}}$ for a small time $t_2=\tau$. Then, we obtain:
\begin{eqnarray} \label{eq:phase}
   {\rm Tr}(\rho^\prime U) &=& {\rm Tr}\left[\frac{\sqrt{\rho_1}}{{\rm Tr}\sqrt{\rho_1}}\left(\mathbb{1}-i\tau \frac{\sqrt{\rho_2}}{{\rm Tr}\sqrt{\rho_2}}\right)\right] \nonumber \\
   &=& \frac{{\rm Tr}\sqrt{\rho_1}}{{\rm Tr}\sqrt{\rho_1}}-i\tau \frac{{\rm Tr}\sqrt{\rho_1}\sqrt{\rho_2}}{{\rm Tr}\sqrt{\rho_1}{\rm Tr}\sqrt{\rho_2}} \nonumber \\
   &=& 1-i\tau \frac{{\rm Tr}\sqrt{\rho_1\rho_2}}{{\rm Tr}\sqrt{\rho_1}{\rm Tr}\sqrt{\rho_2}}.
   \end{eqnarray}
In the above equation ${\rm Tr}\sqrt{\rho_1}\sqrt{\rho_2}$ is written as ${\rm Tr}\sqrt{\rho_1 \rho_2}$ using the fact that $\rho_1$ and $\rho_2$ commute. 
Besides, the visibility $V:=\left|{\rm Tr}(\rho^\prime U)\right|$ must satisfy the relation:    
\begin{eqnarray}\label{eq:visibility}
   V^2 = \left|{\rm Tr}(\rho^\prime U)\right|^2 \nonumber
   &=& \rm{Re}[{\rm Tr}(\rho^\prime U)]^2 + \rm{Im}[{\rm Tr}(\rho^\prime  U)]^2 \\
   &=& 1 + \tau^2\left[\frac{{\rm Tr}\sqrt{\rho_1\rho_2}}{{\rm Tr}\sqrt{\rho_1}{\rm Tr}\sqrt{\rho_2}}\right]^2.
\end{eqnarray}
Specifically, the numerator of the second term in the above equation contains the fidelity between the quantum states $\rho_1$ and $\rho_2$.

\begin{figure}
    \centering
\begin{quantikz}
  \lstick{$\ket{0}$} &\gate{H}&\phase{\phi}&\ctrl{1}&\gate{H}&
    \meter{}&& \rstick{$\sigma$} \\
   \lstick{$\rho^\prime$} &&&\gate{U}&&&&
\end{quantikz}
\caption{Circuit diagram of the Mach-Zehnder interferometer.}
\label{fig:mzi_circuit}
\end{figure}
 Density matrix exponentiation in LMR technique uses the SWAP gate, $S$, that acts on two states as follows: 
\begin{eqnarray}
   S(\rho \otimes \varsigma) S^\dagger &=& \varsigma \otimes \rho,
\end{eqnarray}
so that the LMR algorithm exponentiates the density matrix $\rho$ in a register $A$ to obtain a unitary $e^{i\rho t}$ and gets it to act on some state $\varsigma$ in another register $B$ as follows:
\begin{eqnarray}
   {\rm Tr}_A\left[e^{iSt}(\rho \otimes \varsigma) e^{-iSt}\right] \nonumber
   &=&  {\rm Tr}_A\left[(e^{i\varsigma t}\rho e^{-i\varsigma t}) \otimes (e^{i\rho t}\varsigma e^{-i\rho t})\right] \\ 
   &=& e^{i\rho t}\varsigma e^{-i\rho t}.
\end{eqnarray}
This is achieved by repeatedly doing the below \cite{LMR,KSGS}:
\begin{eqnarray}\label{eq:lmr}
{\rm Tr}_A\left[e^{iS\Delta t}(\rho\otimes\varsigma) e^{-iS\Delta t}\right] = \varsigma - i\Delta t[\rho,\varsigma]+O(\Delta t^2).
\end{eqnarray}
Here, the swap operator $S$ is sparse, which allows $e^{iS\Delta t}$ to be implemented efficiently \cite{BACS,HHL}. Furthermore, the total evolution time $t$ is given by $t=n\Delta t$, where $n=O(t^2/\epsilon)$ is the required number of copies of $\rho$, and consequently the required number of times (\ref{eq:lmr}) must be repeated to simulate $e^{i\rho t}$ with an error of $\epsilon$.

The improved quantum phase estimation (IQPE) method, drawing from Ref.~\cite{HHL}, operates as follows: We begin with two registers in the quantum state $\ket{A_0}\ket{\psi_k}$. Here, $\ket{\psi_k}$  signifies the $k$-th eigenstate of a Hermitian operator, $\Omega$, that we exponentiate. 

The auxiliary state $\ket{A_0}$ is defined as;
$|A_0\rangle:=\sqrt{\frac{2}{T}}\sum_{\iota=0}^{T-1}\sin\frac{\pi(\iota+\frac{1}{2})}{T}|\iota\rangle$, 
where $T$ is a sufficiently large integer. This state $\ket{A_0}$ can be prepared up to an error of $\delta_A$ within a time scaling polynomially with $\log(T/\delta_A)$. Next, we apply a conditional Hamiltonian evolution $\sum_{\iota=0}^{T-1}|\iota\rangle\langle\iota|\otimes e^{i\Omega\iota t/T}$ on the initial state on both the registers. We then perform an inverse quantum Fourier transform on the first register, that yields the state $\sum_{q=0}^{T-1}\upsilon_{q|k}|q\rangle|\psi_k\rangle$. We define $\tilde{\lambda}_q$ as the estimate of the $q$-th eigenvalue $\lambda_q$ of $\Omega$ as $\tilde{\lambda}_q:=\frac{2\pi q}{t}$. This allows us to relabel the Fourier basis states $\ket{q}$ to get the state $\sum_{q=0}^{T-1}\upsilon_{q|k}|\tilde{\lambda}_q\rangle|\psi_k\rangle$. Assuming perfect phase estimation, the coefficient $\upsilon_{q|k}$ would be $1$ only for $\tilde{\lambda}_q=\lambda_k$,
and $0$ otherwise. Thus, we get the state $|\tilde{\lambda}_k\rangle|\psi_k\rangle$, and upon measuring the first register, we get the estimate of $\lambda_k$.

Now, if $\rho=\sum_u \varphi_u \ket{u}\bra{u}$, we obtain using DME scheme, the unitary: $e^{i\rho t}=\sum_u e^{i\varphi_u t}\ket{u}\bra{u}$. The subsequent improved quantum phase estimation (IQPE) process requires controlled-$e^{i\rho t}$ operations $\sum_{\iota=0}^{T-1}|\iota\rangle\langle\iota|\otimes e^{i\rho t\iota/T}$ (for some $T$), that is done by acting the conditional swap operator $|\iota\rangle\langle\iota|\otimes e^{iSt\iota/T}$ on $|\iota\rangle\langle\iota|\otimes\rho\otimes\varsigma$, and tracing out $\rho$. Note that the eigenvalues of $S$ are $\pm 1$, so that $W^{1/T}=W^{2^1/T}=W^{2^2/T}=W^{2^3/T}\ldots$, where $W=e^{iSt}$, if we use say $t=2\pi$. We perform the phase estimation as follows: $\ket{0}\bra{0}\otimes\frac{\mathbb{1}}{2^N} \xrightarrow[e^{i\rho t}]{IQPE} \frac{1}{2^N}\sum_u \ket{\Tilde{\varphi}_u}\bra{\Tilde{\varphi}_u}\otimes\ket{u}\bra{u}$, where $N$ is the number of qubits that constitute the state $\rho$. Taking $z$ copies of ancilla qubits, initialized in the state $\ket{0}$, and rotating it, conditioned on the phase estimates $\tilde{\varphi}_u$, followed by uncomputing the phase estimation (see Ref.~\cite{HHL}), we get:
\begin{eqnarray}\label{eq:ancilla}
     &&\frac{1}{2^N} \sum_u \ket{u}\bra{u}\otimes \left[\sqrt{1-  \frac{\sqrt{\tilde{\varphi}_u}}{\sqrt{\kappa}}}\ket{0}^{\otimes z}+ \sqrt{\frac{\sqrt{\tilde{\varphi}_u}}{\sqrt{\kappa}}}\ket{1}^{\otimes z}\right]\\ \nonumber
    && \times\left[\sqrt{1-\frac{\sqrt{\tilde{\varphi}_u}}{\sqrt{\kappa}}}\bra{0}^{\otimes z}+ \sqrt{\frac{\sqrt{\tilde{\varphi}_u}}{\sqrt{\kappa}}}\bra{1}^{\otimes z}\right].\\ \nonumber
     &&\mbox{The effective state of each of the ancilla qubits will be:}\\ 
     &&\left[(1-\frac{\sum_u \sqrt{\Tilde{\varphi}_u}}{\sqrt{\kappa}})\ket{0}\bra{0}+\frac{\sum_u \sqrt{\Tilde{\varphi}_u}}{\sqrt{\kappa}}\ket{1}\bra{1}\right], 
\end{eqnarray}
which we represent as: $\chi := (1-\lambda)\ket{0}\bra{0}+\lambda\ket{1}\bra{1}$, where $\lambda:=\frac{{\rm Tr}\sqrt{\rho}}{\sqrt{\kappa}}=\sum_u\sqrt{\Tilde{\varphi}_u}/\sqrt{\kappa}$ and $\kappa=\frac{\varphi_{max}}{\varphi_{min}}$ is the condition number of $\rho$. The eigenvalues of the matrix $\rho$ should range between $1/\kappa$ to $1$ or the matrix will be said to be ill-conditioned. To determine $\lambda$, we perform DME using LMR to get the unitary $e^{i\chi t_1}$, and then perform IQPE on the unitary for the eigenstate $\ket{1}$: $\ket{0}\ket{1}\xrightarrow[e^{i\chi t_1}]{IQPE}\ket{\Tilde{\lambda}}\ket{1}$. To perform the DME we need $z=O(t_1^2/\epsilon)$ copies of ancilla qubits. Upon measuring the first register, we obtain the estimate $\Tilde{\lambda}$ of $\lambda$ by multiplying $\sqrt{\kappa}$ with it.
Furthermore, upon post-selecting the state in (\ref{eq:ancilla}) for the ancilla qubit being $\ket{1}$, we get the state $\frac{1}{\lambda}\sum_u \frac{\sqrt{\Tilde{\varphi}_u}}{\sqrt{\kappa}}\ket{u}\bra{u}$, which is an estimate of the state $\sqrt{\rho}/{\rm Tr}\sqrt{\rho}$. This allows us to obtain estimates of $\sqrt{\rho}/{\rm Tr}(\sqrt{\rho})$ and ${\rm Tr}\sqrt{\rho}$ for both $\rho=\rho_1$ and $\rho=\rho_2$.

As illustrated in the quantum circuit of Figure \ref{fig:mzi_circuit} by (\ref{eq:phase}), the phase of the Mach-Zehnder interferometer is directly related to fidelity. 
Beginning with an initial state $(\ket{0}\bra{0} \otimes \rho)$ as input, the output (final) state is obtained as \cite{KPP}: 
\begin{eqnarray}
   \rho_f &=& \frac{1}{2}[\ket{+}\bra{+}\otimes U\rho^\prime U^\dagger + \ket{-} \bra{-}\otimes \rho^\prime + \nonumber\\ 
   && e^{-i\phi}(\ket{+}\bra{-} \otimes U\rho^\prime)+e^{i\phi}(\ket{-}\bra{+}\otimes \rho^\prime U)].
\end{eqnarray}
The effective state of the first qubit in $\rho_f$ above is given by
\begin{eqnarray*}
   \sigma = \frac{1}{2}\left[\ket{+}\bra{+}+\ket{-}\bra{-}+ \alpha e^{-i\phi} \ket{+}\bra{-}+\alpha^\dagger e^{i\phi}\ket{-}\bra{+}\right],
\end{eqnarray*}
where $\alpha := {\rm Tr}(U\rho^\prime)$ and $\alpha^\dagger = {\rm Tr}(\rho^\prime U^\dagger)$. The eigenstates of $\sigma$ are given by
\begin{equation}
   \ket{\chi_+}=\frac{1}{\sqrt{2}}\left[\ket{+}+\ket{-}\right],
\end{equation}
\begin{equation}
   \ket{\chi_-}=\frac{1}{\sqrt{2}}\left[-\ket{+}+\ket{-}\right],
    \end{equation}
    and the eigenvalues of $\sigma$ are $\lambda_+ = 1+ \sqrt{\alpha \alpha^\dagger}=1+V$ and $\lambda_- = 1- \sqrt{\alpha \alpha^\dagger}=1-V$, so that $\sigma$ can be written as follows: 
\begin{equation} \label{eq:sigma}
    \sigma = \lambda_+ \ket{\chi_+}\bra{\chi_+}+ \lambda_- \ket{\chi_-}\bra{\chi_-}.
\end{equation}

We again apply DME to $\sigma$ and perform IQPE to get: $\ket{0}\bra{0}\otimes \frac{\mathbb{1}}{2}\xrightarrow[e^{i\sigma t_3}]{IQPE} \frac{1}{2}[\ket{\Tilde{\lambda}_{+}}\bra{\Tilde{\lambda}_{+}}\otimes \ket{\chi_+}\bra{\chi_+}+ \ket{\Tilde{\lambda}_{-}}\bra{\Tilde{\lambda}_{-}}\otimes \ket{\chi_-}\bra{\chi_-}]$. Through a small number of measurements on the first register, we can readily extract estimates for the eigenvalues, $\Tilde{\lambda}_{+}$ and $\Tilde{\lambda}_{-}$. These estimates, in turn, allow us to determine an estimate for $V=\sqrt{\alpha\alpha^\dagger}$, using which in (\ref{eq:visibility}), we calculate an estimate of the fidelity for two commuting density matrices as ${\rm Tr}\sqrt{\rho_1\rho_2}$.
   
\section{Algorithm} \label{3}
Our algorithm is as follows;
\begin{enumerate}
    \item Implement Density Matrix Exponentiation (DME) on each of the states $\rho=\rho_1$ and $\rho=\rho_2$ to obtain controlled unitaries for Improved Quantum Phase Estimation (IQPE) to be performed in the next step. DME by Lloyd-Mohseni-Rebentrost algorithm (LMR) \cite{LMR} is implemented as follows using Swap operator, $S$ :
    \begin{eqnarray}
   {\rm Tr}_A\left[e^{i S\Delta t}(\rho \otimes \varsigma) e^{-iS\Delta t}\right]  \nonumber
   &=  {\rm Tr}_A\left[(e^{i\varsigma \Delta t}\rho e^{-i\varsigma \Delta t}) \right.\\
   & \left. \otimes  (e^{i\rho \Delta t}\varsigma e^{-i\rho \Delta t})\right] \\ \nonumber 
   &= e^{i\rho \Delta t}\varsigma e^{-i\rho \Delta t},
\end{eqnarray}
which is repeated $n$ number of times to obtain the unitary $e^{i\rho t}$ with $t=n \Delta t$. Upon exponentiating the density matrix of the state $\rho$ in register A, the resulting unitary acts on another state $\varsigma$ in another register B. Here, $\Delta t$ is a short time step. \label{algo:step1}
   \item Perform Improved Quantum Phase Estimation (IQPE) \cite{HHL} with the exponentiated density matrices ($e^{i\rho_1 t}$ and $e^{i\rho_2 t}$) to estimate the eigenvalues of $\rho=\rho_1$ and $\rho=\rho_2$, respectively, as follows: $\ket{0}\bra{0}\otimes\frac{\mathbb{1}}{2^N} \xrightarrow[e^{i\rho t}]{IQPE} \frac{1}{2^N}\sum_u \ket{\Tilde{\varphi}_u}\bra{\Tilde{\varphi}_u}\otimes\ket{u}\bra{u}$, where $N$ is the number of qubits that constitute the state $\rho=\sum_u \varphi_u \ket{u}\bra{u}$ and $\Tilde{\varphi}_u$ is an estimate of $\varphi_u$. \label{algo:step2}
 \item We prepare $z$ number of ancilla qubits in state  $\ket{0}^{\otimes z}$, rotate it based on the phase estimates $\tilde{\varphi}_u$, and then we undo the phase estimation process, to get: 
 \begin{eqnarray} \label{eq:effstate}
 \nonumber 
&\frac{1}{2^N} \sum_u \ket{u}\bra{u}\otimes \left[\sqrt{1-  \frac{\sqrt{\tilde{\varphi}_u}}{\sqrt{\kappa}}}\ket{0}^{\otimes z}+ \sqrt{\frac{\sqrt{\tilde{\varphi}_u}}{\sqrt{\kappa}}}\ket{1}^{\otimes z}\right] \\ 
&\times \left[\sqrt{1-  \frac{\sqrt{\tilde{\varphi}_u}}{\sqrt{\kappa}}}\bra{0}^{\otimes z}+ \sqrt{\frac{\sqrt{\tilde{\varphi}_u}}{\sqrt{\kappa}}}\bra{1}^{\otimes z}\right],
\end{eqnarray}
where $\kappa=\frac{\varphi_{max}}{\varphi_{min}}$ and post-select the above state for an ancilla qubit to be $\ket{1}$, to obtain the state $\frac{1}{\lambda}\sum_u \frac{\sqrt{\Tilde{\varphi}_u}}{\sqrt{\kappa}}\ket{u}\bra{u}$, which is an estimate of the state $\sqrt{\rho}/{\rm Tr}\sqrt{\rho}$,  where $\lambda:=\frac{{\rm Tr}\sqrt{\rho}}{\sqrt{\kappa}}=\sum_u \sqrt{\Tilde{\varphi}_u}/\sqrt{\kappa}$. We do this for both the states $\rho=\rho_1$ and $\rho=\rho_2$. \label{algo:step3}

\item The effective state of each of the ancilla qubits in (\ref{eq:effstate}) is
  $\chi:= (1-\lambda)\ket{0}\bra{0}+\lambda\ket{1}\bra{1}$. We exponentiate $\chi$ and perform phase estimation on the resulting unitary $e^{i \chi t_1}$ for eigenstate $\ket{1}$, where $t_1$ is a small time, to get an estimate $\Tilde{\lambda}$ of $\lambda$, and multiply it with $\sqrt{\kappa}$ to get ${\rm Tr}\sqrt{\rho}$. We do this for both the states $\rho=\rho_1$ and $\rho=\rho_2$. \label{algo:step4}
  \item We compute ${\rm Tr}(\rho_1^{1/2}\rho_2^{1/2})$ using ${\rm Tr}(\rho^\prime U)$, where $\rho^\prime=\sqrt{\rho_1}/{\rm Tr}\sqrt{\rho_1}$, and $U= e^{it_2 \sqrt{\rho_2}/{\rm Tr}\sqrt{\rho_2}}$ (again obtained by DME), for a small time $t_2 =\tau$ where :
  \begin{eqnarray*} 
   {\rm Tr}(\rho^\prime U) 
   = \mathbb{1}-i\tau \frac{{\rm Tr}\sqrt{\rho_1\rho_2}}{{\rm Tr}\sqrt{\rho_1}{\rm Tr}\sqrt{\rho_2}}.
   \end{eqnarray*}
   To accomplish this, DME is performed on $\sigma=  \lambda_+ \ket{\chi_+}\bra{\chi_+}+ \lambda_- \ket{\chi_-}\bra{\chi_-}$ from Fig.~\ref{fig:mzi_circuit} to exponentiate it, followed by IQPE on it to estimate $\lambda_+= 1+ \sqrt{\alpha \alpha^\dagger}$ and $\lambda_-= 1- \sqrt{\alpha \alpha^\dagger}$. Here, $\ket{\chi_+}=\frac{1}{\sqrt{2}}\left[\ket{+}+\ket{-}\right]$, $\ket{\chi_-}=\frac{1}{\sqrt{2}}\left[-\ket{+}+\ket{-}\right]$, $\alpha := {\rm Tr}(U\rho^\prime)$, $\alpha^\dagger = {\rm Tr}(\rho^\prime U^\dagger)$, and $\phi$ is the phase of Mach-Zehnder Interferometer applied as a phase gate. This yields the fidelity ${\rm Tr}(\rho_1^{1/2}\rho_2^{1/2})$ between $\rho_1$ and $\rho_2$ from \cite{akp,KPP}: 
   \begin{eqnarray*}
    \alpha \alpha^\dagger = \left|{\rm Tr}(\rho^\prime U)\right|^2 
   = 1 + \tau^2\left[\frac{{\rm Tr}\sqrt{\rho_1\rho_2}}{{\rm Tr}\sqrt{\rho_1}{\rm Tr}\sqrt{\rho_2}}\right]^2.
   \end{eqnarray*} \label{algo:step5}
\end{enumerate}

\section{Complexity analysis}\label{4}
\begin{enumerate}
    \item The Density Matrix Exponentiation (DME), using the LMR method, has a complexity of:\begin{equation}\label{lmreq}
        O(\log_2 (d)n)= O(Nt^2/\epsilon),
    \end{equation}
    where $d$ is the dimension $2^N$ of $\rho$, and $n=O(t^2/\epsilon)$ is the number of copies of $\rho$ required to simulate $e^{i\rho t}$. 
    Since we do this for both $\rho_1$ and $\rho_2$, the complexity after taking both of them into consideration becomes $O(2Nt^2/\epsilon) \approx O(Nt^2/\epsilon).$ Here, $\epsilon$ is the simulation error. \label{cmplx:step1}
    
    \item Since we perform Improved Quantum Phase Estimation (IQPE) after DME, we will have the time $t=O(\sqrt{\kappa}/\epsilon)$ in (\ref{lmreq}) (see Ref.~\cite{HHL}). Hence, the complexity of DME, followed by IQPE, is $O(\kappa N/\epsilon^3)$, where $\kappa$ is the larger of the condition number of the density matrices.\label{cmplx:step2}
    \item The complexity owing to the controlled rotation operation is subsumed in step \ref{cmplx:step2}.
    \label{cmplx:step3}
    
    \item To obtain $\sqrt{\rho}/{\rm Tr}(\sqrt{\rho})$, we perform a post-selection, the probability of obtaining $|1\rangle$ as measurement outcome for which is $O(1/\kappa)$ (again, see Ref.~\cite{HHL}). The algorithm involves another DME on $\chi$, followed by IQPE, to estimate $\lambda$. The complexity of DME here is $O((\log_2 2)z)$, where $z=O(t_1^2/\epsilon)$ is the required number of copies of $\chi$. Substituting for $t_1=O(1/\epsilon)$, since we do IQPE next, the complexity of this step is $O(1/\epsilon^3)$. This is done for both $\rho_1$ and $\rho_2$. \label{cmplx:step4}

    \item In finding out the quantity ${\rm Tr}\sqrt{\rho_1 \rho_2}$, we need the unitary $U= e^{it_2 \sqrt{\rho_2}/{\rm Tr}\sqrt{\rho_2}}$, for which we need to exponentiate the state $\sqrt{\rho_2}/{\rm Tr}\sqrt{\rho_2}$ obtained earlier. The complexity of DME here will be $O(\log_2(d)w)$ with $w=O(t_2^2/\epsilon)$, where $t_2 = 2$, since there is only $1$ control qubit for the controlled-$U$ operation. Thus, the complexity of this DME is $O(N/\epsilon)$. Afterwards, we perform another DME on $\sigma$ followed by IQPE to estimate the eigenvalues of $\sigma$. 
    This DME will be of complexity $O((\log_22)y))$ with $y=O(t_3 ^2/\epsilon)$, and owing to subsequent IQPE, we have $t_3 = O(1/\epsilon)$. Thus, this step has an overall complexity of $O(N/\epsilon) \times O(1/\epsilon^3) = O(N/\epsilon^4)$. \label{cmplx:step5}
\end{enumerate}
We can then infer the overall complexity of the algorithm as follows. Since the dominating steps \ref{algo:step1} and \ref{algo:step2} are needed to be repeated for every copy of $\sigma$ required in step \ref{algo:step5}, the overall complexity of the algorithm will be $O(\kappa N/\epsilon^3) \times O(N/\epsilon^4) \times O(\kappa) = O(\kappa^2N^2/\epsilon^7)$. Here, $O(\kappa)$ arises from the post-selection probability. 
\section{Numerical Example}\label{5}

To substantiate the theoretical framework and evaluate the dimensional robustness of the proposed fidelity estimation algorithm, we numerically simulate the Mach-Zehnder interferometric protocol across varying system sizes: $N \in \{2, 3, 4\}$ qubits. 

To rigorously test the algorithm, we construct sparse, non-diagonal, commuting mixed density matrices for each dimension.  

The explicitly generated density matrices evaluated in this simulation are defined below. For the 2-qubit system ($d=4$):
\begin{equation*}
\label{eq:rho_1_2Q}
\rho_{1}^{(2Q)} = \begin{bmatrix}
    0.300 & 0.000 & 0.100 & 0.000 \\
    0.000 & 0.200 & 0.000 & 0.100 \\
    0.100 & 0.000 & 0.300 & 0.000 \\
    0.000 & 0.100 & 0.000 & 0.200 \\
\end{bmatrix}
\end{equation*}

\begin{equation*}
\label{eq:rho_2_2Q}
\rho_{2}^{(2Q)} = \begin{bmatrix}
    0.200 & 0.000 & -0.100 & 0.000 \\
    0.000 & 0.300 & 0.000 & -0.100 \\
    -0.100 & 0.000 & 0.200 & 0.000 \\
    0.000 & -0.100 & 0.000 & 0.300 \\
\end{bmatrix}
\end{equation*}

For the 3-qubit system ($d=8$):
\vspace*{-1mm}
    \relscale{0.7}
    \[
    \begin{aligned}
      \rho_{1}^{(3Q)}&=&   \left[\begin{array}{cccccccccccccccc}
    0.225 & 0.000 & 0.000 & 0.000 & 0.175 & 0.000 & 0.000 & 0.000 \\
    0.000 & 0.125 & 0.000 & 0.000 & 0.000 & 0.075 & 0.000 & 0.000 \\
    0.000 & 0.000 & 0.075 & 0.000 & 0.000 & 0.000 & 0.025 & 0.000 \\
    0.000 & 0.000 & 0.000 & 0.075 & 0.000 & 0.000 & 0.000 & 0.025 \\
    0.175 & 0.000 & 0.000 & 0.000 & 0.225 & 0.000 & 0.000 & 0.000 \\
    0.000 & 0.075 & 0.000 & 0.000 & 0.000 & 0.125 & 0.000 & 0.000 \\
    0.000 & 0.000 & 0.025 & 0.000 & 0.000 & 0.000 & 0.075 & 0.000 \\
    0.000 & 0.000 & 0.000 & 0.025 & 0.000 & 0.000 & 0.000 & 0.075 \\
\end{array}\right],
    \end{aligned}
    \]\normalsize

    \relscale{0.6}
    \[
    \begin{aligned}
      \rho_{2}^{(3Q)}&=&   \left[\begin{array}{cccccccccccccccc}
    0.075 & 0.000 & 0.000 & 0.000 & -0.025 & 0.000 & 0.000 & 0.000 \\
    0.000 & 0.075 & 0.000 & 0.000 & 0.000 & -0.025 & 0.000 & 0.000 \\
    0.000 & 0.000 & 0.125 & 0.000 & 0.000 & 0.000 & -0.075 & 0.000 \\
    0.000 & 0.000 & 0.000 & 0.225 & 0.000 & 0.000 & 0.000 & -0.175 \\
    -0.025 & 0.000 & 0.000 & 0.000 & 0.075 & 0.000 & 0.000 & 0.000 \\
    0.000 & -0.025 & 0.000 & 0.000 & 0.000 & 0.075 & 0.000 & 0.000 \\
    0.000 & 0.000 & -0.075 & 0.000 & 0.000 & 0.000 & 0.125 & 0.000 \\
    0.000 & 0.000 & 0.000 & -0.175 & 0.000 & 0.000 & 0.000 & 0.225 \\
\end{array}\right],
    \end{aligned}
    \]\normalsize

For the 4-qubit system ($d=16$), the localized coherence yields the following extended matrices:
\begin{widetext}
\relscale{0.7}
      $\rho_{1}^{(4Q)}=   \left[\begin{array}{cccccccccccccccc}
   0.125 & 0.000 & 0.000 & 0.000 & 0.000 & 0.000 & 0.000 & 0.000 & 0.075 & 0.000 & 0.000 & 0.000 & 0.000 & 0.000 & 0.000 & 0.000 \\
    0.000 & 0.075 & 0.000 & 0.000 & 0.000 & 0.000 & 0.000 & 0.000 & 0.000 & 0.025 & 0.000 & 0.000 & 0.000 & 0.000 & 0.000 & 0.000 \\
    0.000 & 0.000 & 0.075 & 0.000 & 0.000 & 0.000 & 0.000 & 0.000 & 0.000 & 0.000 & 0.025 & 0.000 & 0.000 & 0.000 & 0.000 & 0.000 \\
    0.000 & 0.000 & 0.000 & 0.075 & 0.000 & 0.000 & 0.000 & 0.000 & 0.000 & 0.000 & 0.000 & 0.025 & 0.000 & 0.000 & 0.000 & 0.000 \\
    0.000 & 0.000 & 0.000 & 0.000 & 0.050 & 0.000 & 0.000 & 0.000 & 0.000 & 0.000 & 0.000 & 0.000 & 0.000 & 0.000 & 0.000 & 0.000 \\
    0.000 & 0.000 & 0.000 & 0.000 & 0.000 & 0.050 & 0.000 & 0.000 & 0.000 & 0.000 & 0.000 & 0.000 & 0.000 & 0.000 & 0.000 & 0.000 \\
    0.000 & 0.000 & 0.000 & 0.000 & 0.000 & 0.000 & 0.050 & 0.000 & 0.000 & 0.000 & 0.000 & 0.000 & 0.000 & 0.000 & 0.000 & 0.000 \\
    0.000 & 0.000 & 0.000 & 0.000 & 0.000 & 0.000 & 0.000 & 0.050 & 0.000 & 0.000 & 0.000 & 0.000 & 0.000 & 0.000 & 0.000 & 0.000 \\
    0.075 & 0.000 & 0.000 & 0.000 & 0.000 & 0.000 & 0.000 & 0.000 & 0.125 & 0.000 & 0.000 & 0.000 & 0.000 & 0.000 & 0.000 & 0.000 \\
    0.000 & 0.025 & 0.000 & 0.000 & 0.000 & 0.000 & 0.000 & 0.000 & 0.000 & 0.075 & 0.000 & 0.000 & 0.000 & 0.000 & 0.000 & 0.000 \\
    0.000 & 0.000 & 0.025 & 0.000 & 0.000 & 0.000 & 0.000 & 0.000 & 0.000 & 0.000 & 0.075 & 0.000 & 0.000 & 0.000 & 0.000 & 0.000 \\
    0.000 & 0.000 & 0.000 & 0.025 & 0.000 & 0.000 & 0.000 & 0.000 & 0.000 & 0.000 & 0.000 & 0.075 & 0.000 & 0.000 & 0.000 & 0.000 \\
    0.000 & 0.000 & 0.000 & 0.000 & 0.000 & 0.000 & 0.000 & 0.000 & 0.000 & 0.000 & 0.000 & 0.000 & 0.050 & 0.000 & 0.000 & 0.000 \\
    0.000 & 0.000 & 0.000 & 0.000 & 0.000 & 0.000 & 0.000 & 0.000 & 0.000 & 0.000 & 0.000 & 0.000 & 0.000 & 0.050 & 0.000 & 0.000 \\
    0.000 & 0.000 & 0.000 & 0.000 & 0.000 & 0.000 & 0.000 & 0.000 & 0.000 & 0.000 & 0.000 & 0.000 & 0.000 & 0.000 & 0.050 & 0.000 \\
    0.000 & 0.000 & 0.000 & 0.000 & 0.000 & 0.000 & 0.000 & 0.000 & 0.000 & 0.000 & 0.000 & 0.000 & 0.000 & 0.000 & 0.000 & 0.050 \\
\end{array}\right]$,
\end{widetext}

\begin{widetext}
\relscale{0.7}
     $\rho_{2}^{(4Q)}=  \left[\begin{array}{cccccccccccccccc}
    0.050 & 0.000 & 0.000 & 0.000 & 0.000 & 0.000 & 0.000 & 0.000 & -0.025 & 0.000 & 0.000 & 0.000 & 0.000 & 0.000 & 0.000 & 0.000 \\
    0.000 & 0.050 & 0.000 & 0.000 & 0.000 & 0.000 & 0.000 & 0.000 & 0.000 & -0.025 & 0.000 & 0.000 & 0.000 & 0.000 & 0.000 & 0.000 \\
    0.000 & 0.000 & 0.050 & 0.000 & 0.000 & 0.000 & 0.000 & 0.000 & 0.000 & 0.000 & -0.025 & 0.000 & 0.000 & 0.000 & 0.000 & 0.000 \\
    0.000 & 0.000 & 0.000 & 0.050 & 0.000 & 0.000 & 0.000 & 0.000 & 0.000 & 0.000 & 0.000 & -0.025 & 0.000 & 0.000 & 0.000 & 0.000 \\
    0.000 & 0.000 & 0.000 & 0.000 & 0.075 & 0.000 & 0.000 & 0.000 & 0.000 & 0.000 & 0.000 & 0.000 & -0.025 & 0.000 & 0.000 & 0.000 \\
    0.000 & 0.000 & 0.000 & 0.000 & 0.000 & 0.075 & 0.000 & 0.000 & 0.000 & 0.000 & 0.000 & 0.000 & 0.000 & -0.025 & 0.000 & 0.000 \\
    0.000 & 0.000 & 0.000 & 0.000 & 0.000 & 0.000 & 0.075 & 0.000 & 0.000 & 0.000 & 0.000 & 0.000 & 0.000 & 0.000 & -0.025 & 0.000 \\
    0.000 & 0.000 & 0.000 & 0.000 & 0.000 & 0.000 & 0.000 & 0.125 & 0.000 & 0.000 & 0.000 & 0.000 & 0.000 & 0.000 & 0.000 & -0.075 \\
    -0.025 & 0.000 & 0.000 & 0.000 & 0.000 & 0.000 & 0.000 & 0.000 & 0.050 & 0.000 & 0.000 & 0.000 & 0.000 & 0.000 & 0.000 & 0.000 \\
    0.000 & -0.025 & 0.000 & 0.000 & 0.000 & 0.000 & 0.000 & 0.000 & 0.000 & 0.050 & 0.000 & 0.000 & 0.000 & 0.000 & 0.000 & 0.000 \\
    0.000 & 0.000 & -0.025 & 0.000 & 0.000 & 0.000 & 0.000 & 0.000 & 0.000 & 0.000 & 0.050 & 0.000 & 0.000 & 0.000 & 0.000 & 0.000 \\
    0.000 & 0.000 & 0.000 & -0.025 & 0.000 & 0.000 & 0.000 & 0.000 & 0.000 & 0.000 & 0.000 & 0.050 & 0.000 & 0.000 & 0.000 & 0.000 \\
    0.000 & 0.000 & 0.000 & 0.000 & -0.025 & 0.000 & 0.000 & 0.000 & 0.000 & 0.000 & 0.000 & 0.000 & 0.075 & 0.000 & 0.000 & 0.000 \\
    0.000 & 0.000 & 0.000 & 0.000 & 0.000 & -0.025 & 0.000 & 0.000 & 0.000 & 0.000 & 0.000 & 0.000 & 0.000 & 0.075 & 0.000 & 0.000 \\
    0.000 & 0.000 & 0.000 & 0.000 & 0.000 & 0.000 & -0.025 & 0.000 & 0.000 & 0.000 & 0.000 & 0.000 & 0.000 & 0.000 & 0.075 & 0.000 \\
    0.000 & 0.000 & 0.000 & 0.000 & 0.000 & 0.000 & 0.000 & -0.075 & 0.000 & 0.000 & 0.000 & 0.000 & 0.000 & 0.000 & 0.000 & 0.125 \\
\end{array}\right]$
 \end{widetext}

To determine the physical boundary where the required first-order Taylor approximation of the controlled unitary ($U \approx \mathbb{I} - i\tau\frac{\sqrt{\rho_2}}{\text{Tr}\sqrt{\rho_2}}$) fails, we subjected all three systems to an extended interaction parameter sweep, $\tau \in [0.01, 1.0]$.

\begin{figure}[htbp]
    \centering
    \includegraphics[width=1\linewidth]{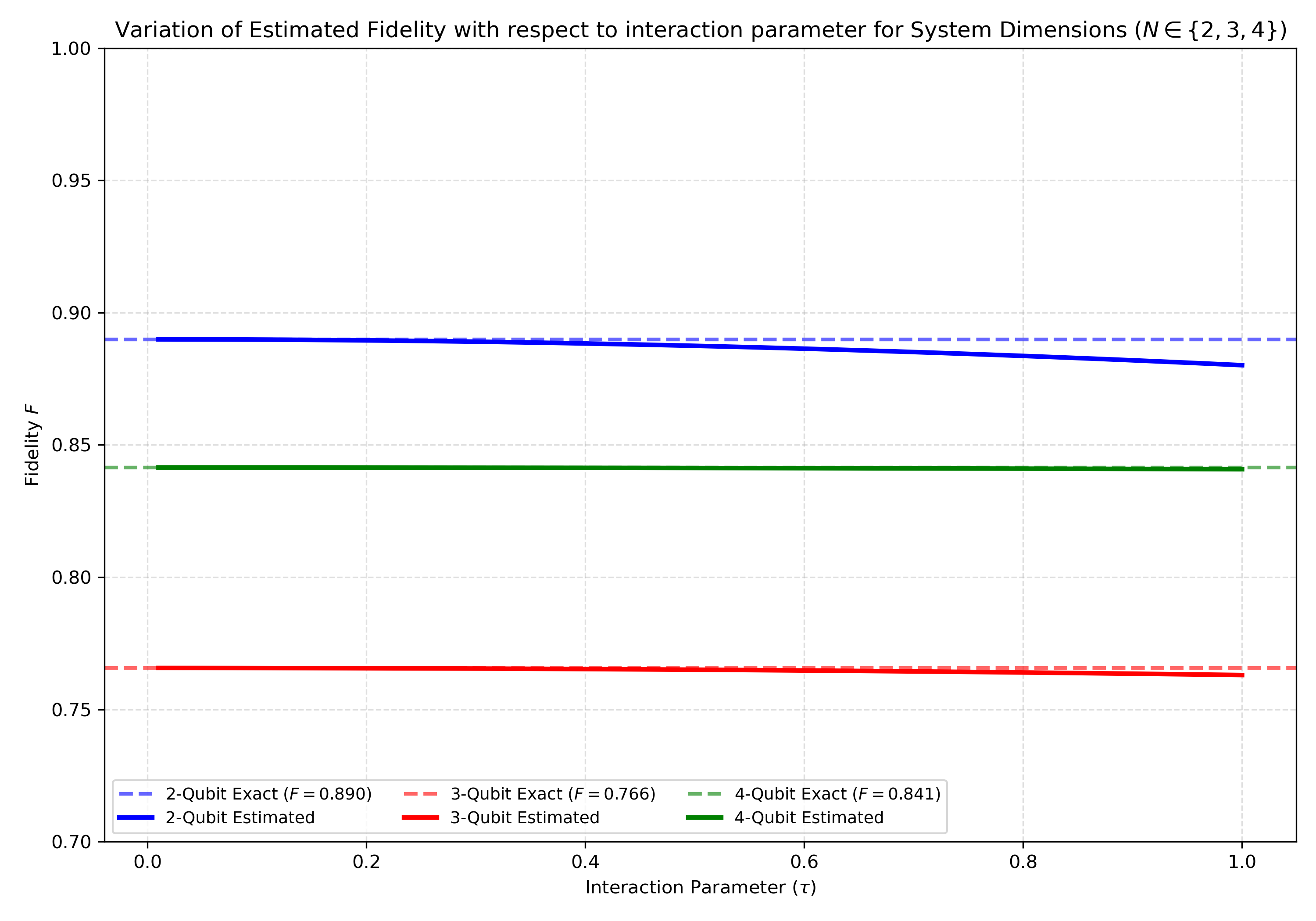}
    \caption{Algorithm approximation breakdown across system dimensions ($N \in \{2, 3, 4\}$). The estimated fidelity exactly tracks the analytical Uhlmann fidelity (dashed lines) within the strictly perturbative regime ($\tau \ll 1$). The nonlinear divergence point, governed by the uncompensated $\mathcal{O}(\tau^2)$ error term, scales favorably with the Hilbert space dimension.}
    \label{fig:tau_multidim}
\end{figure}

As illustrated in Fig.~\ref{fig:tau_multidim}, the algorithm successfully isolates the exact theoretical fidelity for all system dimensions provided $\tau$ is sufficiently small. However, as the interaction time is forcefully extended, the higher-order terms in the exponentiation—specifically the uncompensated second-order $\mathcal{O}(\tau^2)$ term—induce a measurable divergence from the analytical target.

Crucially, the simulation reveals a highly advantageous dimensional scaling effect: the interaction parameter threshold at which the linear approximation breaks down is pushed significantly higher as the number of qubits increases. The 2-qubit estimation diverges rapidly for $\tau > 0.1$, while the 4-qubit estimation remains structurally stable approaching $\tau \approx 1.0$. 

This behavior is mathematically governed by the trace normalization factor located in the denominator of the exponentiated generator, $K = \frac{\sqrt{\rho_2}}{\text{Tr}\sqrt{\rho_2}}$. Because the trace of the square root of a density matrix scales proportionally with the Hilbert space dimension ($\text{Tr}\sqrt{\rho_2} \le \sqrt{d} = 2^{N/2}$), the overall spectral norm of $K$ strictly decreases for higher-dimensional systems. Consequently, the uncompensated error term $\mathcal{O}(\tau^2 K^2)$ is heavily suppressed in larger Hilbert spaces. This intrinsic "dimensional dilution" grants high-dimensional states a significantly wider operational margin for the physical interaction parameter $\tau$, empirically reinforcing the scalability of the interferometric extraction technique for complex quantum systems.

\section{Discussion}\label{6}

Estimating the fidelity between two unknown density operators, $\rho$ and $\sigma$, highlights the severe limitations of standard Quantum State Tomography (QST). Because QST relies on full state estimation to reconstruct the entire density matrix classically before computing the trace overlap, its sample and measurement complexity scales exponentially with the number of qubits $N$, making it completely impractical for systems larger than a few qubits. In stark contrast, our new algorithm circumvents full reconstruction by targeting only the necessary underlying properties. This shifts the task from an intractable global reconstruction to an efficient, scalable property-estimation problem. Therefore, it is instructive to qualitatively contrast our approach with the conventional route to fidelity estimation via full quantum state tomography \cite{NC}. Standard tomographic reconstruction of an unknown density matrix requires measuring a complete set of observables sufficient to fix all $d^2-1$ real parameters. For a $d$-dimensional system ($d=2^N$), the number of measurement settings and consequently, the number of state copies needed for faithful reconstruction scales exponentially with the number of qubits $N$ for generic states. Furthermore, classical post-processing via the standard linear-inversion and maximum-likelihood schemes \cite{James2001, Hradil1997} scale as $O(d^3)$ or worse in the system dimension \cite{Smolin2012}, with measurement resources growing exponentially.

Crucially, tomography solves a strictly harder problem than necessary: it recovers the entire density matrix, whereas the Uhlmann fidelity is merely a single scalar functional. Our algorithm sidesteps this exponential overhead by bypassing quantum state tomography. Instead, it extracts only the necessary scalar quantities such as $\text{Tr}\sqrt{\rho_1\rho_2}$, the normalization factors $\text{Tr}\sqrt{\rho_1}$ and $\text{Tr}\sqrt{\rho_2}$ via density matrix exponentiation and phase estimation acting directly on the unknown states. 
Consequently, the resource cost of our scheme scales polynomially in $N$ (specifically as $\mathcal{O}(\kappa^2N^2/\epsilon^7)$), rather than exponentially. We explicitly note that this scaling trade-off introduces an additional dependence on the condition number $\kappa$ and a comparatively steep inverse polynomial dependence on the algorithmic precision $\epsilon$. Nevertheless, this makes our protocol highly preferable whenever the end goal is fidelity estimation alone, particularly for higher-dimensional systems where tomographic overhead becomes strictly prohibitive. It should be emphasized that this favorable polynomial scaling is specific to the commuting case and provides a highly efficient, targeted alternative to full state characterization.

The error paramater $\epsilon$ is an error in trace distance for simulating exponentials of a density matrix, and is twice the error in trace distance in improved quantum phase estimation precision. An error in trace distance determines the maximum possible probability of error, since the trace distance between two states gives the maximum difference in probability of any measurement on the two states \cite{KLLOY}. Thus, the parameter $\epsilon$ determines the maximum probability of simulation error and twice the maximum probability of estimation precision error. Usually, this quantity $\epsilon$ is needed to be exponentially as small as $O(1/2^N)$ to capture distinctly all the eigenvalues of a density matrix for a state of $N$ qubits. However, if we do not require to capture all the eigenvalues distinctly, a polynomially as small as $O(1/{\rm poly}(N))$ value for the quantity $\epsilon$ may often suffice (along with using $T=O(2^N)$ in improved quantum phase estimation), even for high-rank density matrices being used, as long as the overall cumulative probability of error of our algorithm is less than $1/3$. In that case, our algorithm complexity would be $O({\rm poly}(N))$, which would imply that our algorithm would be efficient. This is why our algorithm may serve as a resource-efficient technique to compute the fidelity between two quantum states commuting with each other, even with high-rank density matrices. We add a note here that our algorithm will not be efficient, if $\kappa$ is too large, and our algorithm will fail, if $\kappa$ is too small and too close to $1$, since then the controlled rotation will not happen.

\section{Conclusion}\label{7}
In summary, we have presented a new quantum algorithm for the resource-efficient estimation of fidelity between two unknown general quantum states commuting with each other, encompassing both pure and mixed states. Our approach bypasses the demanding process of full quantum state tomography by leveraging the Lloyd-Mohseni-Rebentrost technique for density matrix exponentiation and integrating it with improved quantum phase estimation, augmented with interference of mixed state, to extract eigenvalues. Subsequent controlled rotations and post-selection yield the normalized square roots of the density matrices. Crucially, employing the Mach-Zehnder interferometer for mixed state directly enables the calculation of fidelity. This protocol significantly reduces the experimental overhead typically associated with fidelity determination, offering a scalable and practical solution for benchmarking quantum devices, validating quantum operations, and advancing the characterization of complex quantum systems in the noisy intermediate-scale quantum (NISQ) era and beyond. 
Our algorithm has an overall time complexity of $O(\kappa^2N^2/\epsilon^7)$, where $N$ is the number of qubits in each of the two given states, and $\epsilon$ is a precision error. In cases, where $\epsilon$ is not necessarily taken as exponentially small, but only polynomially small, as long as the overall probability of error of our algorithm is below $1/3$, our algorithm would have a time complexity of $O({\rm poly}(N))$, and so would be efficient, even for high-rank density matrices.

\section*{Conflict of Interest}
The authors declare no conflict of interest.

\medskip

\bibliographystyle{unsrt}
\bibliography{fidelity}

\end{document}